\documentclass[aps, pra, reprint]{revtex4-2}
\usepackage{graphicx}
\usepackage{amsmath, amssymb, amsfonts}
\usepackage{dsfont}
\usepackage{bm}

\begin{document}

\title{Hilbert space fragmentation from lattice geometry}

\author{Pieter H. Harkema}
\thanks{These authors contributed equally to this work.}
\affiliation{Department of Physics and Astronomy, Aarhus University, DK-8000 Aarhus C, Denmark}

\author{Michael Iversen}
\thanks{These authors contributed equally to this work.}
\affiliation{Department of Physics and Astronomy, Aarhus University, DK-8000 Aarhus C, Denmark}

\author{Anne E. B. Nielsen}
\affiliation{Department of Physics and Astronomy, Aarhus University, DK-8000 Aarhus C, Denmark}

\begin{abstract}
The eigenstate thermalization hypothesis describes how isolated many-body quantum systems reach thermal equilibrium.
However, quantum many-body scars and Hilbert space fragmentation violate this hypothesis and cause nonthermal behavior.
We demonstrate that Hilbert space fragmentation may arise from lattice geometry in a spin-$1/2$ model that conserves the number of domain walls.
We generalize a known, one-dimensional, scarred model to larger dimensions and show that this model displays Hilbert space fragmentation on the Vicsek fractal lattice and the two-dimensional lattice.
Using Monte Carlo methods, the model is characterized as strongly fragmented on the Vicsek fractal lattice when the number of domain walls is either small or close to the maximal value.
On the two-dimensional lattice, the model is strongly fragmented when the density of domain walls is low and weakly fragmented when the density of domain walls is high.
Furthermore, we show that the fragmentation persists at a finite density of domain walls in the thermodynamic limit for the Vicsek fractal lattice and the two-dimensional lattice.
We also show that the model displays signatures similar to Hilbert space fragmentation on a section of the second-generation hexaflake fractal lattice and a modified two-dimensional lattice.
We study the autocorrelation function of local observables and demonstrate that the model displays nonthermal dynamics.
\end{abstract}

\maketitle

\section{Introduction}
Isolated, many-body, quantum systems are typically thermal, and the microcanonical ensemble accurately describes the expectation values of local observables at long times.
The eigenstate thermalization hypothesis (ETH) explains this behavior from an ansatz about the matrix elements of local observables \cite{Srednicki1994, Deutsch1998, Rigol2008}.
The ETH has been verified by numerous works (see Ref.~\cite{Luca2016} and references therein).
However, several mechanisms violate the ETH and, as a consequence, cause nonthermal behavior.

Scarred models host a small number of ETH-violating energy eigenstates, called quantum many-body scars (QMBSs), embedded in an otherwise thermal spectrum \cite{Serbyn2021, Moudgalya2022, Chandran2023}.
While scar states are nonthermal, they only represent a weak violation of the ETH.
QMBSs date back to the discovery of analytic excited energy eigenstates in the Affleck-Kennedy-Lieb-Tasaki model \cite{Arovas1989, Moudgalya2018_a, Moudgalya2018_b}, and signatures of QMBSs were observed in experiments with kinetically constrained Rydberg atoms \cite{Bernien2017}.
Since the initial findings, numerous scarred models have been discovered, e.g., Refs.~\cite{Iadecola2019, Schecter2019, Iadecola2019, Shibata2020, Mark2020Aug, Moudgalya2020Aug, Iadecola2020, Chertkov2021, Wildeboer2021}, and QMBSs have been realized in various experimental setups \cite{Bluvstein2021, Chen2022, Zhang2023, Su2023, Zhou2023, Gustafson2023}.

Hilbert space fragmentation (HSF) represents another ETH-violating phenomenon where the Hilbert space is separated into dynamically disconnected subspaces even after resolving all symmetries \cite{Moudgalya2022}.
These subspaces are called Krylov subspaces, and the number of subspaces grows exponentially with system size.
The subspaces may vary in size from one-dimensional ``frozen configurations'' to subspaces with exponentially large dimensions.
While the Krylov subspaces are not described by conventional quantum numbers associated with a symmetry of the Hamiltonian operator, they may be labeled by ``statistically localized integrals of motion'' \cite{Rakovszky2020} or commutant algebras \cite{Moudgalya2022Mar}.
HSF may represent a weak or strong violation of the ETH, and the corresponding model is denoted as, respectively, weakly or strongly fragmented \cite{Sala2020}.
When the largest Krylov subspace constitutes a vanishingly small fraction of the relevant symmetry sector in the thermodynamic limit, the model is strongly fragmented.
On the other hand, the model is weakly fragmented when the size of a Krylov subspace converges to the dimension of its symmetry sector in the thermodynamic limit.
Although HSF violates the ETH, the Krylov subspaces may thermalize according to the Krylov-restricted ETH \cite{Moudgalya2021}.
They may also be integrable or many-body localized \cite{Tomasi2019, Herviou2021}.
HSF has been extensively studied in models with charge and dipole conservation \cite{Pai2019, Morningstar2020, Sala2020, Khemani2020, Moudgalya2021}, but fragmentation also arises in other settings \cite{Yang2020, Lee2021, Hahn2021, Langlett2021, Mukherjee2021Aug, Mukherjee2021Oct, Li2021, Bastianello2022, Richter2022, Pietro2023}.
HSF is linked to Stark many-body localization where the effective Hamiltonian conserves both the charge and dipole moment \cite{Taylor2020, Doggen2021}.
Consequently, signatures of HSF have been observed in experiments with the tilted Fermi-Hubbard model \cite{Scherg2021, Kohlert2022}.

The nature of fragmented models depends on the system dimension.
For instance, the \mbox{$t$ {\textendash} $J_z$} model displays HSF in one dimension~\cite{Rakovszky2020}, but it is not fragmented in two dimensions~\cite{Moudgalya2022Mar}.
Likewise, while the pair-flip model displays HSF in one dimension, it is scarred in dimensions larger than one~\cite{Caha2018, Moudgalya2022Mar}.
Hence, the presence of HSF in one dimension does not guarantee its existence in higher dimensions.
Several works have, however, constructed models that are HSF in dimensions larger than one~\cite{Khemani2020, Yoshinaga2022, Hart2022, Alexey2022, Anwesha2023}.
These findings call for a better understanding of the connection between HSF and the number of spatial dimensions.
Besides analyzing fragmentation in integer dimensions, a general analysis would naturally study HSF on fractal lattices.
Fractal lattices have a more complicated structure, and one may wonder whether HSF can arise from the lattice geometry itself.
If this is possible, then lattices may be constructed where HSF is present on certain sublattices but not on other sublattices.

In this work, we generalize a known, domain-wall conserving, one-dimensional, scarred model~\cite{Iadecola2020} to lattices of dimensions larger than one.
While the one-dimensional model hosts a few nonthermal scar states, the Hilbert space shatters into numerous dynamically disconnected subspaces when the model is placed in higher dimensions.
We study the model on the Vicsek fractal lattice, the two-dimensional lattice, a section of the second-generation hexaflake fractal lattice, and a modified two-dimensional lattice.
We show that the model displays HSF on the Vicsek fractal lattice and the two-dimensional lattice.
For the Vicsek fractal lattice, the fragmentation is strong in symmetry sectors where the number of domain walls is either close to zero or close to the maximal number of domain walls.
For the two-dimensional lattice, the model is strongly fragmented when the density of domain walls is low and weakly fragmented when the density of domain walls is high.
We also show that the fragmentation persists at a finite domain wall density in the thermodynamic limit on the Vicsek fractal lattice and the two-dimensional lattice.
We observe features similar to HSF on a section of the second-generation hexaflake fractal lattice and the modified two-dimensional lattice.
Furthermore, we find that the level of fragmentation depends on the lattice geometry.
This result is explained by lattice sites with more than two nearest neighbors confining the movement of domain walls.
The combination of domain wall conservation and conservation of total magnetization was previously shown to cause HSF in a one-dimensional system~\cite{Yang2020}.
Furthermore, domain wall conservation has been shown to cause HSF on the two-dimensional lattice with periodic boundary conditions~\cite{Yoshinaga2022, Hart2022}.
We show that domain wall conservation also causes HSF on other lattices with dimensions larger than one.
We demonstrate the nonthermal nature of the model by studying the time-averaged autocorrelation function of a local observable, and we compare the results with the Mazur bound.

In Sec.~\ref{sec:model}, we take a one-dimensional, domain-wall conserving, scarred model as our starting point and generalize the model to larger dimensional lattices.
We present the Vicsek fractal lattice, the hexaflake fractal lattice, the two-dimensional lattice, and a modified two-dimensional lattice.
In Sec.~\ref{sec:fragmentation}, we demonstrate that the model displays HSF on the Vicsek fractal lattice and the two-dimensional lattice by explicitly constructing an exponential number of Krylov subspaces for both lattices.
Furthermore, we observe that the model displays characteristics similar to HSF on a section of the second-generation hexaflake fractal lattice and a modified two-dimensional lattice.
We describe the mechanisms that restrict the movement of domain walls and cause fragmentation.
For the Vicsek fractal lattice, we estimate the dimensions of the symmetry sectors with a small number of domain walls using Monte Carlo importance sampling.
We also estimate the dimension of the largest Krylov subspace in each symmetry sector.
These results show that the largest Krylov subspace represents a vanishingly small part of the full symmetry sector.
Hence, the model is strongly fragmented on the Vicsek fractal lattice for a small number of domain walls.
We extend this result to symmetry sectors where the number of domain walls is close to the maximal value.
For the two-dimensional lattice, we compute the dimensions of the Krylov subspaces and symmetry sectors exactly for various system sizes.
Based on this data, we characterize the model as strongly fragmented when the density of domain walls is low and weakly fragmented when the density of domain walls is high.
In Sec.~\ref{sec:autocorrelationfunctions}, we study the long-time average of the autocorrelation function of a local observable and compare the results with the Mazur bound.
While the time-averaged autocorrelation function does not converge to the Mazur bound for the considered lattices, the Mazur bound becomes tight when the Hamiltonian is perturbed by a block diagonal random matrix.
In Sec.~\ref{sec:conclusion}, we summarize the results.

\section{Model}\label{sec:model}
We take the model from Ref.~\cite{Iadecola2020} as our starting point.
Consider a one-dimensional lattice of length $N$ with open boundary conditions described by the Hamiltonian
\begin{equation}\label{eq:1dhamiltonian}
    H_\text{1D} = \lambda\sum_{i = 2}^{N-1} (\sigma_i^x - \sigma^{z}_{i-1}\sigma^{x}_{i}\sigma^{z}_{i+1}) + \Delta\sum_{i = 1}^{N} \sigma_{i}^{z} + J\sum_{i = 1}^{N-1}\sigma_{i}^{z}\sigma_{i+1}^{z},
\end{equation}
where $\sigma_{i}^{x}$ and $\sigma_{i}^{z}$ are the Pauli $x$- and $z$-operators acting on site $i$.
The kinetic term $\sigma_i^x - \sigma^{z}_{i-1}\sigma^{x}_{i}\sigma^{z}_{i+1}$ flips the spin on site $i$ if the sum of the spins on the two nearest neighbors, sites $i - 1$ and $i + 1$, is zero.
The second term in Eq.~\eqref{eq:1dhamiltonian} represents an energy contribution ascribed to the spin orientation in a uniform magnetic field in the $z$-direction with strength $\Delta$.
The third term represents nearest neighbor interactions along the $z$-direction with strength $J$.
When two nearest neighbor spins are in different states, the edge connecting the two sites is denoted as a domain wall.
The Hamiltonian operator conserves the number of domain walls.
This model has been shown to host two towers of QMBSs \cite{Iadecola2020}.

We generalize the one-dimensional model from Eq.~\eqref{eq:1dhamiltonian} to higher dimensional lattices while preserving its characteristic features.
In particular, the model should conserve the number of domain walls.
We consider the Hamiltonian
\begin{equation}
    H = H_\lambda + H_z + H_{zz}
    \label{eq:generalized-Hamiltonian}
\end{equation}
with
\begin{subequations}
\begin{align}
    H_\lambda &= \lambda \sum_{\bm r} \delta\Big(\sum_{\langle \bm r, {\bm r}' \rangle} \sigma_{\bm r'}^z \Big) \sigma_{\bm r}^x,
    \label{eq:generalized-Hamiltonian-kinetic-term}\\
    H_z &= \Delta \sum_{\bm r} \sigma_{\bm r}^z,
    \label{eq:generalized-Hamilonian-magnetic-field-interaction}\\
    H_{zz} &= J \sum_{\langle \bm r, \bm r' \rangle} \sigma_{\bm r}^z \sigma_{\bm r'}^z,
    \label{eq:generalized-Hamiltonian-nearest-neighbor-interaction}
\end{align}\label{eq:generalized-Hamiltonian-terms}%
\end{subequations}
where $\langle \cdot, \cdot \rangle$ refers to nearest neighbor sites, and $\delta$ is the function given by $\delta(0) = 1$ and $\delta(x) = 0$ for $x \neq 0$.
The three-body kinetic terms in Eq.~\eqref{eq:1dhamiltonian} are, hence, generalized to the operator $H_\lambda$ consisting of many-body terms.
The remaining terms in Eqs.~\eqref{eq:generalized-Hamilonian-magnetic-field-interaction} and \eqref{eq:generalized-Hamiltonian-nearest-neighbor-interaction} are similar to the one-dimensional model.
We note that Eqs.~\eqref{eq:generalized-Hamiltonian} and \eqref{eq:generalized-Hamiltonian-terms} reduce to Eq.~\eqref{eq:1dhamiltonian} for the one-dimensional lattice.

\begin{figure*}
    \centering
    \includegraphics{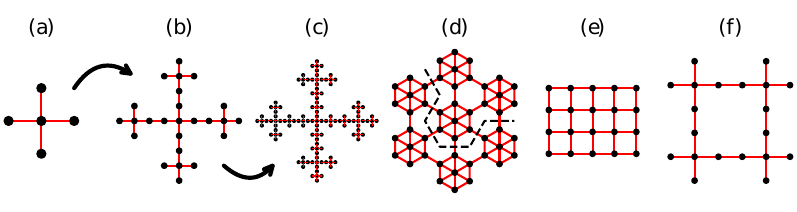}
    \caption{
    Illustration of the considered lattices.
    Black dots show the lattice sites and red lines display nearest neighbor edges.
    (a) The first-generation Vicsek fractal lattice.
    (b) The second-generation Vicsek fractal lattice is obtained from generation one by substituting all lattice sites with five new sites.
    (c) Similarly, the third-generation Vicsek fractal lattice is obtained from generation two by substituting all sites with generation one Vicsek fractal lattices.
    (d) The second-generation hexaflake fractal lattice.
    We consider the lattice consisting of three connected first-generation hexaflake fractal lattices (above the dashed line).
    (e) The two-dimensional lattice of size $L_x \times L_y = 5 \times 4$.
    (f) The modified two-dimensional lattice constructed from four connected first-generation Vicsek fractal lattices.
}
    \label{fig:Lattices}
\end{figure*}
\begin{figure}
    \centering
    \includegraphics{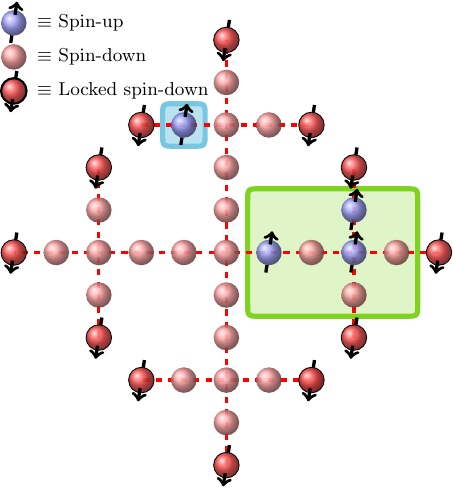}
    \caption{
    The second-generation Vicsek fractal lattice padded with dynamically inactive sites along the boundary (dark red balls with black outlines and downward pointing arrows).
    The figure illustrates a product state in the symmetry sector with $n_\text{dw} = 8$ domain walls.
    Four sites are spin-up (blue balls with upward pointing arrows) and the remaining sites are spin-down (red balls).
    The sites inside the green box are denoted as an ``active arm''.
    These sites may flip their spin and the domain walls may move around within the sublattice.
    Note, however, that the spin-down site connected to the active arm from the left (red ball at the center of the lattice) can not change its state to spin-up because three or more of its nearest neighbors are always spin-down.
    Therefore, this site effectively locks the dynamics within the active arm and all spin-down sites outside the green box can not be flipped.
    Similarly, the spin-up site inside the blue box is dynamically inactive and can not be flipped.
    Product states where all spins are dynamically inactive are denoted as ``frozen states'', and we construct an exponential number of such states in Appendix~\ref{appendix:Hilbert-space-fragmentation-proof}.}
    \label{fig:vicsekfractal}
\end{figure}

Equation~\eqref{eq:generalized-Hamiltonian-kinetic-term} implies that sites with an odd number of nearest neighbors display no dynamics.
Therefore, we generally study lattices where all sites not on the lattice boundary have an even number of nearest neighbors.
Furthermore, we introduce additional sites along the boundary such that boundary sites with an odd number of nearest neighbors get an even number of nearest neighbors.
We hereby ensure that all sites in the original lattice are dynamically active.
The newly added sites are inactive and we generally choose them as spin-down.
We consider open boundary conditions for all lattices throughout this work.
We study the Vicsek fractal lattice, a section of the second-generation hexaflake fractal lattice, the two-dimensional lattice, and a modified two-dimensional lattice.
Figure~\ref{fig:Lattices} displays the considered lattices.
The figure also illustrates how the Vicsek fractal lattice of generation $g$ is obtained from generation $g - 1$ by substituting all lattice sites with five new sites.
All studied lattices have boundary sites with an odd number of nearest neighbors and thus get additional sites appended.
Figure~\ref{fig:vicsekfractal} illustrates the procedure of adding inactive sites along the boundary for the second-generation Vicsek fractal lattice.

\section{Hilbert space fragmentation}\label{sec:fragmentation}
Fragmented models are characterized by the Hilbert space separating into kinetically disconnected subspaces
\begin{subequations}
\begin{align}
    \mathcal{H} &= \bigoplus_{i} \mathcal{K}_i, \\
    \mathcal{K}_i &= \text{span}\big(\{ H^n |\psi_i\rangle | n=0, 1, 2, \ldots \} ),
\end{align}\label{eq:krylovspaces}%
\end{subequations}
where $\mathcal H$ is the full Hilbert space, $\mathcal{K}_i$ denotes a Krylov subspace of dimension $d_i$, and $\lvert \psi_i \rangle$ is a product state that generates $\mathcal{K}_i$.
Let $|s_1 \ldots s_N\rangle$ be a simultaneous eigenket of the spin operators in the $z$-direction, i.e., $\{\sigma_{\bm r}^z / 2\}$ where $\sigma_{\bm r}^z$ is the Pauli $z$-operator on site $\bm r$.
The Krylov subspace generated by a state $\lvert s_1 \ldots s_N \rangle$ may be computed by iterative look-up in the matrix representation of the kinetic part of the Hamiltonian $H_{\lambda}$.
In essence, if $\langle s_1' \ldots s_N' \vert H_\lambda \vert s_1 \ldots s_N \rangle \neq 0$ then $\vert s_1' \ldots s_N' \rangle$ belong to the same Krylov subspace as $\vert s_1 \ldots s_N \rangle$.
Since the Hamiltonian operator conserves the number of domain walls, the Krylov subspaces may be constructed by considering each symmetry sector separately.

\begin{figure*}
    \centering
    \includegraphics{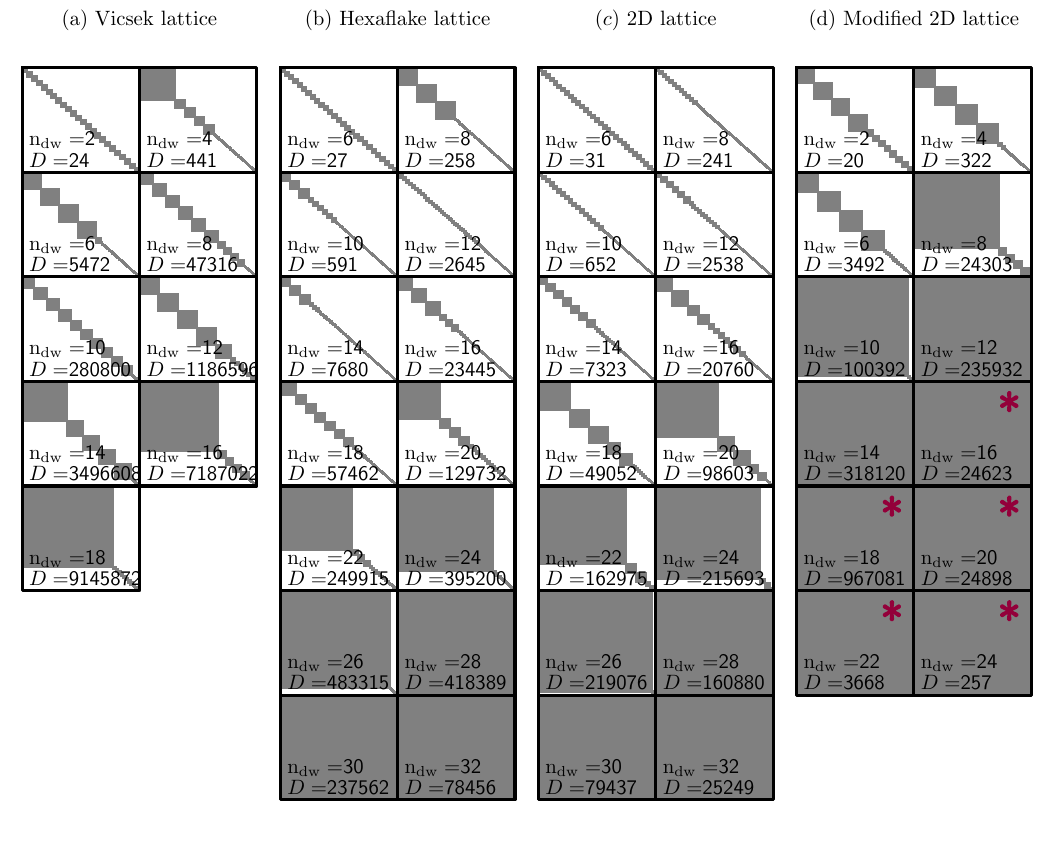}
    \caption{
    The Hamiltonian in each symmetry sector for (a) the second-generation Vicsek fractal lattice illustrated in Fig.~\ref{fig:Lattices}(b) with $25$ dynamically active sites, (b) the section of the second-generation hexaflake fractal lattice illustrated in Fig.~\ref{fig:Lattices}(d) with $21$ dynamically active sites, (c) the two-dimensional lattice illustrated in Fig.~\ref{fig:Lattices}(e) of size $L_x \times L_y = 5 \times 4$, and (d) the modified two-dimensional lattice illustrated in Fig.~\ref{fig:Lattices}(f) with $20$ dynamically active sites.
    Gray pixels represent nonzero matrix elements and white pixels correspond to vanishing matrix elements.
    The figure illustrates the block diagonal structure of the Hamiltonian operator due to Hilbert space fragmentation.
    $n_\text{dw}$ denotes the number of domain walls characterizing the symmetry sector, and $D \equiv D_{n_\text{dw}}$ denotes the dimension of the symmetry sector.
    For the Vicsek fractal lattice, the Hamiltonian is only depicted up to $n_\text{dw}=18$ domain walls since the Krylov subspaces in the symmetry sector with $n_\text{dw}$ domain walls have the same sizes as the Krylov subspaces in the symmetry sector with $n_\text{dw}^\text{max} - n_\text{dw}$ domain walls as discussed in Appendix~\ref{appendix:domainw-wall-symmetry}.
    The red asterisks mark symmetry sectors containing just a single Krylov subspace that spans the full sector.
    }
    \label{fig:fragmentation}
\end{figure*}

In this section, we demonstrate that the model from Eq.~\eqref{eq:generalized-Hamiltonian} displays HSF on certain lattices and we characterize the fragmentation in various symmetry sectors.
However, before delving into this analysis, it is instructive to inspect the fragmentation of the model for a single system size of the considered lattices.
We consider the second-generation Vicsek fractal lattice with $25$ dynamically active sites from Fig.~\ref{fig:Lattices}(b), the section of the second-generation hexaflake fractal lattice with $21$ dynamically active sites from Fig.~\ref{fig:Lattices}(d), the two-dimensional lattice of size $L_x \times L_y = 5 \times 4$ from Fig.~\ref{fig:Lattices}(e), and the modified two-dimensional lattice with $20$ dynamically active sites from Fig.~\ref{fig:Lattices}(f).
The system sizes are chosen such that the considered lattices consist of approximately the same number of sites.
Figure~\ref{fig:fragmentation} displays the matrix representation of the Hamiltonian operator in the basis $\{|s_1 \ldots s_N \rangle\}$ for the single system size of each lattice.
We show each symmetry sector separately, and the basis states are arranged to allow the block diagonal structure from Eq.~\eqref{eq:krylovspaces}.
The gray pixels represent nonzero matrix elements while white pixels represent vanishing matrix elements.
For all considered lattices, we generally observe that the Hilbert space shatters into numerous Krylov subspaces.
However, the degree of fragmentation depends on the number of domain walls and the geometry of the lattice.

\subsection{The Vicsek fractal lattice}
\label{sec:Vicsek_fractal_lattice}
The kinetic term in the Hamiltonian operator $H_\lambda$ on the Vicsek fractal lattice is invariant under the transformation that inserts domain walls on edges where there is no domain wall and removes domain walls from edges where there is a domain wall.
Therefore, the size and number of Krylov subspaces in the sector with $n_\text{dw}$ domain walls are identical to the size and number of the Krylov subspaces in the sector with $n_\text{dw}^\text{max} - n_\text{dw}$ domain walls, where $n_\text{dw}^\text{max}$ is the maximal number of domain walls on the Vicsek fractal lattice of generation $g$.
Without loss of generality, we study sectors with less than or equal to half filling of domain walls, i.e., $n_\text{dw} \leq n_\text{dw}^\text{max} / 2$.
We further discuss this point in Appendix \ref{appendix:domainw-wall-symmetry}.

Figure~\ref{fig:fragmentation}(a) displays the Hamiltonian operator for the second-generation Vicsek fractal lattice consisting of $25$ dynamically active sites.
We observe fragmentation within all symmetry sectors, with the amount of fragmentation varying between sectors.
The fragmentation is caused by lattice sites with more than two nearest neighbors which may restrict the movement of domain walls.
For instance, if all domain walls are located within one ``arm'' of the Vicsek fractal lattice, then the central spin can not be flipped, and it will act as a blockade for the movement of the domain walls as illustrated in Fig.~\ref{fig:vicsekfractal}.
The dynamics of each Krylov subspace is, therefore, restricted to isolated regions on the lattice, with the remaining sites being frozen.
Following this line of thought, we prove in Appendix~\ref{appendix:Hilbert-space-fragmentation-proof} that the model displays HSF on the Vicsek fractal lattice.
In the proof, we construct an exponential number of one-dimensional Krylov subspaces by trapping the domain walls in certain regions of the lattice.
We also show that the fragmentation persists at finite domain wall densities in the thermodynamic limit.

Figure \ref{fig:fragmentation}(a) illustrates that symmetry sectors with few domain walls seem to consist of many small Krylov subspaces.
On the other hand, sectors with the number of domain walls close to $n_\text{dw}^\text{max} / 2$ seem dominated by one large Krylov subspace.
Hence, the Vicsek fractal lattice seems to display stronger fragmentation for fewer domain walls when $n_\text{dw} \leq n_\text{dw}^\text{max} / 2$.
We characterize the model as either weakly or strongly fragmented by studying the largest Krylov subspace in each symmetry sector.
Let $\mathcal{H}_{n_\text{dw}}$ be the symmetry sector with $n_\text{dw}$ domain walls, and let $D_{n_\text{dw}}$ be the dimension of this sector.
Furthermore, let $d^\text{max}_{n_\text{dw}}$ be the dimension of the largest Krylov subspace in the sector with $n_\text{dw}$ domain walls.
The model is strongly fragmented if $\lim_{g\to \infty}(d_{n_\text{dw}}^\text{max}/D_{n_\text{dw}}) = 0$ and weakly fragmented if $\lim_{g\to \infty} (d_{n_\text{dw}}^\text{max}/ D_{n_\text{dw}}) = 1$.
However, an exact calculation of $D_{n_\text{dw}}$ and $d_{n_\text{dw}}^\text{max}$ is only feasible for small system sizes, i.e., for generations $g \leq 2$.
For larger generations, we utilize Monte Carlo importance sampling to estimate these quantities.
We outline the numerical procedure in Appendix \ref{appendix:monte_carlo_methods}.
Figure \ref{fig:krylov_subspace_size} displays the ratio $d_{n_\text{dw}}^\text{max}/D_{n_\text{dw}}$ as a function of the lattice generation $g$ for different symmetry sectors $n_\text{dw}$.
The symmetry sector with no domain walls $n_\text{dw} = 0$ only consists of the state with all spins down $\lvert \downarrow \downarrow \ldots \downarrow \rangle$.
Therefore, the ratio is unity for all generations.
For all other considered sectors, the ratio $d_{n_\text{dw}}^\text{max}/D_{n_\text{dw}}$ decreases with increasing generation.
These results indicate that the largest Krylov subspace represents a vanishing small part of the full symmetry sector.
Therefore, the system displays strong fragmentation for sectors with a small number of domain walls.
We remark that the domain wall density of the considered symmetry sectors with $n_\text{dw} \in \{0, 2, \ldots, 10\}$ domain walls goes to zero when the generation is taken to infinity.
Hence, the numerical results in Fig.~\ref{fig:krylov_subspace_size} do not characterize the fragmentation at nonzero domain wall densities in the thermodynamic limit.

The above analysis concerns symmetry sectors with less than or equal to half-filling of domain walls $n_\text{dw} \leq n_\text{dw}^\text{max} / 2$.
Recall that $d_{n_\text{dw}}^\text{max} = d_{n_\text{dw}'}^\text{max}$ and $D_{n_\text{dw}} = D_{n_\text{dw}'}$ with $n_\text{dw}' = n_\text{dw}^\text{max} - n_\text{dw}$.
Therefore, the system also displays strong fragmentation in symmetry sectors where the number of domain walls is close to being maximal.

\begin{figure}
    \centering
    \includegraphics{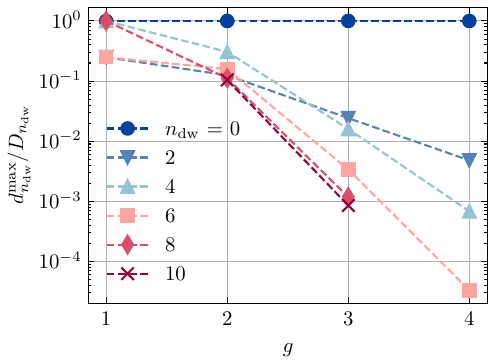}
    \caption{
    The ratio between the dimension of the largest Krylov subspace $d_{n_\text{dw}}^\text{max}$ and the dimension of the symmetry sector $D_{n_\text{dw}}$ as a function of the generation $g$ of the Vicsek fractal lattice.
    Each graph corresponds to a fixed number of domain walls $n_\text{dw}$.
    The data for generation $g = 1, 2$ is exact while the data for generation $g = 3, 4$ is obtained using Monte Carlo importance sampling as described in Appendix \ref{appendix:monte_carlo_methods}.
    The symmetry sector with $n_\text{dw} = 0$ domain walls is one-dimensional for all generations, and the ratio is unity.
    For all other considered sectors, the ratio decreases with increasing generation indicating that the system is strongly fragmented.
    }
    \label{fig:krylov_subspace_size}
\end{figure}

\subsection{The two-dimensional lattice}
\begin{figure}
    \includegraphics[width=\linewidth]{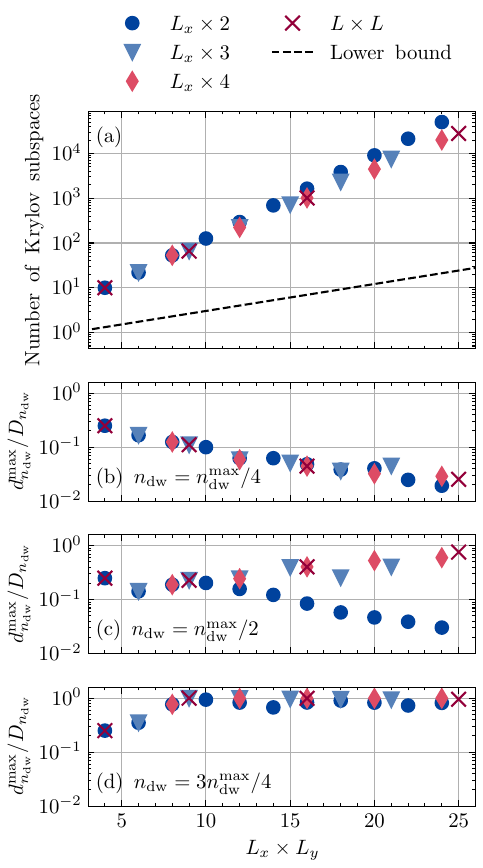}
    \caption{
        (a) The number of Krylov subspaces as a function of system size $L_x \times L_y$.
        We vary the length of one side of the lattice $L_x$ while keeping the other side fixed at $L_y = 2$ (dark blue dots), $L_y = 3$ (light blue triangles), and $L_y = 4$ (light red diamonds).
        We also increase the length $L$ of both sides simultaneously $L_x \times L_y = L \times L$ (dark red crosses).
        Notice that some data points are included in more than one of these groups and have multiple markers, e.g., $L_x \times L_y = 2 \times 2$.
        The number of Krylov subspaces scales exponentially with system size and is larger than the lower bound (dashed line) obtained in Appendix~\ref{appendix:two_dimensional_lattice_fragmentation}.
        (b)-(d) the ratio between the dimension of the largest Krylov subspace and the dimension of the corresponding symmetry sector as a function of system size for various densities of domain walls.
        We consider (b) one-quarter filling $n_\text{dw} = n_\text{dw}^\text{max} / 4$, (c) half-filling $n_\text{dw} = n_\text{dw}^\text{max} / 2$, and (d) three-quarter filling $n_\text{dw} = 3 n_\text{dw}^\text{max} / 4$.
        The results indicate that the model is strongly fragmented for a low density of domain walls and weakly fragmented for a high density of domain walls.
    }
    \label{fig:square_lattice}
\end{figure}
Next, we study the model from Eq.~\eqref{eq:generalized-Hamiltonian} on the two-dimensional lattice of size $L_x \times L_y = 5 \times 4$.
Figure~\ref{fig:fragmentation}(c) displays the Hamiltonian operator in each symmetry sector as a block diagonal matrix.
The figure shows that symmetry sectors with a small number of domain walls shatter into numerous Krylov subspaces and the figure hints that the model is fragmented on the two-dimensional lattice.
We formally prove this statement in Appendix~\ref{appendix:two_dimensional_lattice_fragmentation} by computing a lower bound on the number of Krylov subspaces.
We show that this lower bound scales exponentially with system size, and hence, that the model displays HSF on the two-dimensional lattice.
We also demonstrate that the fragmentation persists at a finite density of domain walls in the thermodynamic limit.
Figure~\ref{fig:square_lattice}(a) shows the number of Krylov subspaces for various system sizes $L_x \times L_y$ and the lower bound from Appendix~\ref{appendix:two_dimensional_lattice_fragmentation}.
The figure illustrates that the number of Krylov subspaces scales exponentially with system size and that the number of Krylov subspaces grows faster with system size than the lower bound.

We characterize the fragmentation of the model on the two-dimensional lattice by computing the Krylov subspaces in certain symmetry sectors for different system sizes.
Similarly to the analysis in Sec.~\ref{sec:Vicsek_fractal_lattice}, we study the ratio between the dimension $d_{n_\text{dw}}^\text{max}$ of the largest Krylov subspace in the symmetry sector with $n_\text{dw}$ domain walls and the dimension $D_{n_\text{dw}}$ of the corresponding symmetry sector.
The system size is varied by extending one side of the lattice while keeping the length of the other side fixed, e.g., the system sizes $L_x \times 4$ for different values of $L_x$.
We also vary the system size by increasing the length of both sides simultaneously, i.e., the square lattices $L_x \times L_y = L \times L$ for different values of $L$.
Figures~\ref{fig:square_lattice}(b)-(d) display the ratio $d_{n_\text{dw}}^\text{max} / D_{n_\text{dw}}$ as a function of system size.
We consider symmetry sectors with a specific filling of domain walls, e.g., half-filling $n_\text{dw} = n_\text{dw}^\text{max} / 2$, and we round to the nearest valid number of domain walls when necessary.
For one-quarter filling $n_\text{dw} = n_\text{dw}^\text{max} / 4$, we find that the ratio $d_{n_\text{dw}}^\text{max} / D_{n_\text{dw}}$ decreases with increasing system size both when increasing the length of one side and when increasing both sides simultaneously.
This result indicates that the model displays strong fragmentation on the two-dimensional lattice for a low density of domain walls.
At half-filling $n_\text{dw} = n_\text{dw}^\text{max} / 2$, the ratio decreases for system sizes $L_x \times 2$ when increasing $L_x$.
However, the ratio increases for the lattices $L_x \times 3$ and $L_x \times 4$ when increasing $L_x$ and for the square lattices $L \times L$ when increasing $L$.
At three-quarter filling $n_\text{dw} = 3 n_\text{dw}^\text{max} / 4$, the ratio increases and approaches unity when increasing the length of one side or both sides.
This result indicates that the model is weakly fragmented at a high density of domain walls.
Recall that sites with four or more nearest neighbors can restrict the movement of domain walls by acting as blockades.
On the two-dimensional lattice, all sites have four nearest neighbors and every site may, therefore, limit the movement of the domain walls. 
Indeed, we find that the model is strongly fragmented in symmetry sectors with a low density of domain walls.
However, we observe that the model is weakly fragmented in symmetry sectors with a high density of domain walls.
We interpret this result as sites failing to act as blockades when the density of domain walls is sufficiently large.

We remark that a model similar to Eq.~\eqref{eq:generalized-Hamiltonian} was previously studied on the two-dimensional lattice with periodic boundary conditions~\cite{Yoshinaga2022, Hart2022}.
This model was proven to display HSF when both side lengths of the lattice are a multiple of an odd integer larger than one~\cite{Yoshinaga2022}.
This argument also applies to our model, but for completeness, we provide the proof in Appendix~\ref{appendix:two_dimensional_lattice_fragmentation} which is valid for all system sizes.
The model with periodic boundary conditions was also shown to display strong fragmentation for a low density of domain walls and weak fragmentation for a high density of domain walls~\cite{Hart2022}.
However, the presence of respectively weak and strong fragmentation for periodic boundary conditions does not guarantee the same fragmentation strength for open boundary conditions.
Indeed, the domain walls are more confined for open boundary conditions and one might naively expect that the model will display stronger fragmentation in this case.
Interestingly, we observe that the fragmentation displays the same behavior as a function of domain wall density for open boundary conditions as for periodic boundary conditions.

\subsection{The hexaflake fractal lattice and the modified two-dimensional lattice}
Finally, we consider the model on the section of the second-generation hexaflake fractal lattice with $21$ dynamically active sites from Fig.~\ref{fig:Lattices}(d) and the modified two-dimensional lattice with $20$ dynamically active sites from Fig.~\ref{fig:Lattices}(f).
Figure~\ref{fig:fragmentation}(b) displays the Krylov subspaces in various symmetry sectors for the section of the second-generation hexaflake fractal lattice.
All sites in this lattice have four or six nearest neighbors and, similarly to the two-dimensional lattice, every site may act as a blockade to the domain walls.
We observe somewhat similar fragmentation on the section of the second-generation hexaflake fractal lattice as on the two-dimensional lattice of size $L_x \times L_y = 5 \times 4$.
In particular, we observe that the model on the hexaflake fractal lattice is more fragmented in symmetry sectors with fewer domain walls as compared to symmetry sectors with many domain walls.

The modified two-dimensional lattice is structurally similar to the Vicsek fractal lattice as it is constructed from four first-generation Vicsek fractal lattices.
It consists of sites with four nearest neighbors which may act as blockades and sites with two nearest neighbors that do not restrict the movement of the domain walls.
However, in contrast to the Vicsek fractal lattice, the modified two-dimensional lattice introduces a ``loop''.
Figure~\ref{fig:fragmentation}(d) illustrates the Krylov subspaces for each symmetry sector on the modified two-dimensional lattice.
We observe that the model displays a smaller amount of fragmentation on this lattice as compared to the second-generation Vicsek fractal lattice.
In particular, symmetry sectors with $n_\text{dw} \geq 16$ domain walls display no fragmentation.
The reduced amount of fragmentation may be related to the presence of a loop in the modified two-dimension lattice.
Similarly, the presence of loops in the two-dimensional lattice and hexaflake fractal lattice may be related to the reduced fragmentation of the model on these lattices in symmetry sectors with a large number of domain walls.

We emphasize that the results presented in this section concern a single system size for the hexaflake fractal lattice and the modified two-dimensional lattice and the observed behavior may not generalize to larger system sizes.

\section{The autocorrelation function of local observables}
\label{sec:autocorrelationfunctions}
The autocorrelation function of a local observable is an effective tool for characterizing systems exhibiting HSF \cite{Sala2020, Rakovszky2020, Moudgalya2022, Moudgalya2022Mar}.
For a local operator $\mathcal O$ acting within a Hilbert space of dimension $\mathcal D$, the infinite temperature autocorrelation function at time $t$ is given by
\begin{align}
    \mathcal{A}(t) = \langle {\mathcal O}(t) {\mathcal O}(0) \rangle = \frac{1}{\mathcal D}\text{Tr}\big(e^{i{H}t} {\mathcal O} e^{-i{H}t} {\mathcal O} \big),
    \label{eq:autocorrelation}
\end{align}
where $\langle \cdot \rangle = \mathrm{Tr}(\cdot) / \mathcal D$ is the infinite temperature expectation value and $H$ is the Hamiltonian operator governing the system.
Fragmented models are characterized by parts of the system acting as blockades and thereby restricting the movement within the system.
A local observable is, therefore, correlated with itself at later times because the full system is not explored freely.
In contrast, thermal systems do not retain local information about their initial state and the autocorrelation functions of local observables vanish.
These considerations are formally captured by the Mazur inequality which provides a lower bound on the long-time average of the autocorrelation function \cite{Mazur1969, Suzuku1971, Dhar2021}.
The time-averaged autocorrelation function is given by
\begin{align}
    \bar{\mathcal{A}}(T) = \frac{1}{T} \int_0^T \mathcal{A}(t) \, \mathrm{d}t.
\end{align}
Consider a model described by the Hamiltonian $H$ with a set of conserved quantities $\{I_i\}$, i.e., $[I_i, H] = 0$.
The Mazur bound is then given by
\begin{align}
    M_{\mathcal O} = \sum_{ij} \langle \mathcal O^\dagger I_i \rangle [C^{-1}]_{ij} \langle I_j^\dagger \mathcal O \rangle,
    \label{eq:general-mazur-bound}
\end{align}
with $C_{ij} = \langle I_i^\dagger I_j \rangle$.
The time-averaged autocorrelation function satisfies the Mazur inequality
\begin{equation}
    M_{\mathcal O} \leq \lim_{T\to \infty} \bar{\mathcal A}(T).
    \label{eq:Mazur-inequality}
\end{equation}
For a thermal system, the Mazur bound is close to zero, and the infinite time average of the autocorrelation function of generic observables vanishes.
On the other hand, the Mazur bound for integrable systems may be significantly different from zero, and the infinite time averaged autocorrelation function does not relax to zero.
The autocorrelation function may, therefore, identify nonthermal behavior.

Equation~\eqref{eq:general-mazur-bound} may be simplified by diagonalizing $C_{ij}$, i.e., choosing $\{I_i\}$ such that $\langle I_i^\dagger I_j \rangle = \delta_{ij}$.
Then the Mazur bound reduces to $M_{\mathcal O} = \sum_i |\langle I_i^\dagger \mathcal O \rangle|^2$.
Equation~\eqref{eq:Mazur-inequality} is trivially satisfied when the set of conserved quantities is taken as the projections onto the energy eigenstates, i.e., $I_i \propto \vert E_i \rangle \langle E_i \vert$ where $\{\vert E_i \rangle\}$ is the set of energy eigenstates.
However, the Mazur bound may be finite with only a few terms $\vert \langle I_i^\dagger \mathcal O \rangle |^2$ contributing significantly to $M_{\mathcal O}$.
In this case, the finite autocorrelation of $\mathcal O$ may be attributed directly to the conserved quantities contributing significantly to $M_{\mathcal O}$.

For fragmented models, the projection operators onto the Krylov subspaces $\{ {\mathcal P}_i \}$ is a set of conserved quantities.
The projection operators satisfy $\langle \mathcal P_i^\dagger \mathcal P_j\rangle = \delta_{ij} d_i / D_{n_\text{dw}}$ where $d_i$ is the dimension of the $i$-th Krylov subspace, $D_{n_\text{dw}}$ is the dimension of the symmetry sector with $n_\text{dw}$ domain walls, and $\delta_{ij}$ is the Kronecker delta.
We consider the observable $\tilde s_{\bm r}^z = s_{\bm r}^z - \langle s_{\bm r}^z \rangle$ related to the spin-$1/2$ operator $s_{\bm r}^z = \sigma_{\bm r}^z / 2$ at position $\bm r$.
The Mazur bound is given by
\begin{align}
    M_{\tilde s^z_{\bm r}}
    = \frac{1}{D_{n_\text{dw}}}\sum_{i} \frac{\text{Tr}({\mathcal P}_i \tilde s^z_{\bm r})^2}{d_i}.
    \label{eq:mazur-bound}
\end{align}

\begin{figure}
    \centering
    \includegraphics{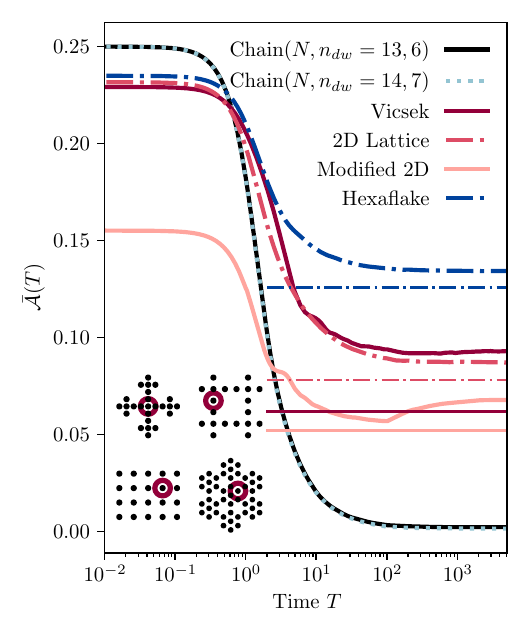}
    \caption{
    The time average of the autocorrelation function for the four considered lattices and two symmetry sectors on the one-dimensional lattice of $N$ sites.
    We consider the symmetry sector $n_\text{dw} = 6$ $(D_{n_\text{dw}} = 5472)$ for the second-generation Vicsek fractal lattice, $n_\text{dw} = 14$ $(D_{n_\text{dw}} = 7680)$ for the section of the second-generation hexaflake fractal lattice, $n_\text{dw} = 14$ $(D_{n_\text{dw}} = 5323)$ for the two-dimensional lattice, and $n_\text{dw} = 6$ $(D_{n_\text{dw}} = 3492)$ for the modified two-dimensional lattice.
    We consider the autocorrelation function of the operator $\tilde s_{\bm r}^z = s_{\bm r}^z - \langle s_{\bm r}^z \rangle$ for the site $\bm r$ illustrated in the inset (inside red circle).
    We find similar results for other sites.
    In all cases, we consider parameters $\lambda = J = 1$ and $\Delta = 0.1$.
    The horizontal lines show the Mazur bound obtained from the projection operators onto the Krylov subspaces.
    }
    \label{fig:autocorrelation}
\end{figure}

We compute the time-averaged autocorrelation function for the one-dimensional lattice, the second-generation Vicsek fractal lattice, a section of the second-generation hexaflake fractal lattice, the two-dimensional lattice, and the modified two-dimensional lattice.
We study two symmetry sectors for the one-dimensional lattice and a single symmetry sector for all other considered lattices.
We consider parameter values $\lambda = J = 1$ and $\Delta = 0.1$ in all cases.
Figure~\ref{fig:autocorrelation} displays the time-averaged autocorrelation function and the corresponding Mazur bound.
In the one-dimensional, scarred model, the time-averaged autocorrelation function converges to a value close to zero.
This behavior is expected for a model where the majority of energy eigenstates are thermal.
On the four lattices with dimensions larger than one, the Mazur bound is larger than zero, and, consequently, the autocorrelation function converges to a finite value.
Hence, the model displays nonthermal behavior on these lattices and the system retains some memory of its initial state.
We remark that the Mazur bound is not tight for the considered lattices with dimensions larger than one.
However, in appendix~\ref{appendix:perturbation}, we show that the Mazur bound becomes tight when the Hamiltonian is perturbed in each Krylov subspace by a random matrix drawn from the Gaussian orthogonal ensemble.

\section{Conclusion}\label{sec:conclusion}
We generalized a known, one-dimensional, scarred model to higher dimensional lattices.
The model displays HSF on the Vicsek fractal lattice and the two-dimensional lattice, with the Hilbert space shattering into an exponential number of one-dimensional Krylov subspaces for both lattices.
For the Vicsek fractal lattice, we demonstrated that the largest Krylov subspace constitutes a vanishingly small fraction of the symmetry sector when the number of domain walls is close to zero or close to being maximal.
Therefore, the model is strongly fragmented on the Vicsek fractal lattice in these symmetry sectors.
For the two-dimensional lattice, the model is strongly fragmented for a low density of domain walls and weakly fragmented for a high density of domain walls. 
For the Vicsek fractal lattice and the two-dimensional lattice, we also showed that the fragmentation persists at a finite domain wall density in the thermodynamic limit.
The model displays features similar to HSF on a section of the second-generation hexaflake fractal lattice and a modified two-dimensional lattice.
We studied the time-averaged autocorrelation function of the $z$-component of a single spin and compared it with the corresponding Mazur bound.
We demonstrated that the model displays nonthermal dynamics on all the considered lattices with dimensions larger than one by observing a finite value of the long-time average of the autocorrelation function.
We also showed that the Mazur bound becomes tight when the Hamiltonian is perturbed by a block diagonal random matrix.
This work demonstrates that Hilbert space fragmentation may arise solely from the conservation of the number of domain walls in dimensions larger than one and that the nature of the fragmentation depends on the lattice geometry.

\begin{acknowledgments}
This work was supported by Carlsbergfondet under Grant No.~CF20-0658 and by Danmarks Frie Forskningsfond under Grant No.~8049-00074B.
\end{acknowledgments}

\appendix
\section{Domain wall - domain wall absence symmetry}
\label{appendix:domainw-wall-symmetry}
The Vicsek fractal lattice is bipartite and we denote the two parts by $A$ and $B$.
Let $A$ be the part containing all the dynamically active sites on the boundary of the Vicsek fractal lattice.
Recall that we introduced additional spin-down sites along the boundary of the Vicsek fractal lattice in Sec.~\ref{sec:model} to ensure all sites in the original lattice have an even number of nearest neighbors.
We note that part $A$ does not contain any of these dynamically inactive sites.
We consider the operator
\begin{align}
    T = \bigotimes_{\bm r \in A} \sigma_{\bm r}^x,
    \label{eq:domain-wall-domain-wall-hole-transformation}
\end{align}
where $\sigma_{\bm r}^x$ is the Pauli $x$-operator acting on site $\bm r$.
Recall that a domain wall is an edge between two nearest neighbor sites of opposite spin orientations along the $z$-direction, i.e., $\uparrow\downarrow$ or $\downarrow\uparrow$.
We consider a lattice edge to be empty if no domain wall is present and an edge to be occupied if a domain wall is present.
The operator $T$ removes all domain walls from occupied edges and inserts domain walls on empty edges.
One may show by direct calculation that $T$ commutes with the kinetic part of the Hamiltonian for any generation of the Vicsek fractal lattice $[T, H_\lambda] = 0$.
Note, however, that the remaining terms in the Hamiltonian operator do not commute with $T$.
The sizes of the Krylov subspaces in the sector with $n_\text{dw}$ domain walls are identical to the sizes of the Krylov subspaces in the sector with $n_\text{dw}^\text{max} - n_\text{dw}$ domain walls where $n_\text{dw}^\text{max}$ is the maximal number of domain walls.
To illustrate this point, consider a state $|\psi\rangle$ with $n_\text{dw}$ domain walls generating the Krylov subspace $\mathcal{K}$.
The corresponding state $|\psi'\rangle = T|\psi \rangle$ obtained by removing domain walls from occupied edges and inserting domain walls on empty edges has $n_\text{dw}^\text{max} - n_\text{dw}$ domain walls.
This state generates a different Krylov subspace $\mathcal{K}'$.
Since $T$ represents a one-to-one correspondence between $\mathcal K$ and $\mathcal K'$, the Krylov subspaces have identical dimensions $\mathrm{dim}(\mathcal{K}) = \mathrm{dim}(\mathcal{K}')$.
Using this fact, we may, without loss of generality, only consider symmetry sectors with the number of domain walls $n_\text{dw} \leq n_\text{dw}^\text{max} / 2$ on the Vicsek fractal lattice.

We remark that the hexaflake fractal lattice, the two-dimensional lattice, and the modified two-dimensional lattice are also bipartite and each lattice may be separated into two parts $A$ and $B$.
However, for these lattices, both parts $A$ and $B$ contain some of the dynamically inactive, spin-down sites introduced as padding to the lattices in Sec.~\ref{sec:model}.
Consequently, the kinetic part of the Hamiltonian is not invariant under $T$, and we generally consider all symmetry sectors for these lattices.

\section{Construction of an exponential number of Krylov subspaces on the Vicsek fractal lattice}
\label{appendix:Hilbert-space-fragmentation-proof}
\begin{figure}
    \centering
    \includegraphics{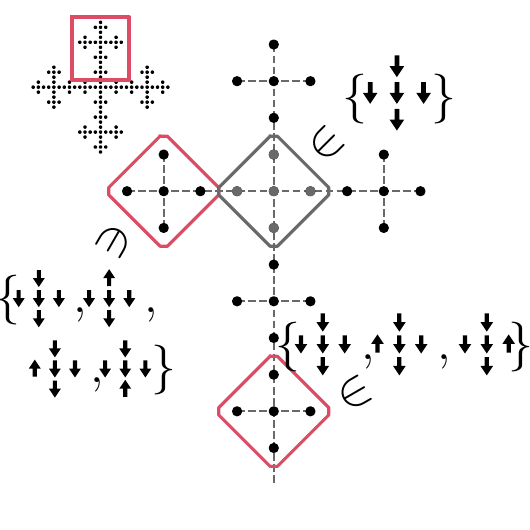}
    \caption{
    A part of the Vicsek fractal lattice of generation $g = 3$.
    We consider the lattice to consist of generation one fractal lattices.
    Some generation one fractals are nearest neighbors with four other generation one fractals (gray sites).
    We choose all sites to be spin-down in these generation one fractals.
    For generation one fractals with two nearest neighbors, the spins are chosen among the three frozen configurations shown in the figure.
    For generation one fractals with one nearest neighbor, we choose the spins among the four frozen configurations shown in the figure.
    We construct an exponential number of eigenstates of the Hamiltonian as the tensor product of these frozen configurations on the generation one fractals.
    }
    \label{fig:Vicsek-fractal-generation-3-colored}
\end{figure}
We prove that the model in Eq.~\eqref{eq:generalized-Hamiltonian} displays Hilbert space fragmentation on the Vicsek fractal lattice by explicitly constructing a set of Krylov subspaces.
We focus on one-dimensional Krylov subspaces and demonstrate that the number of such subspaces scales exponentially with the system size.
We also show that the fragmentation persists at a finite domain wall density in the thermodynamic limit.

Consider the Vicsek fractal lattice of generation one.
There exist several ``frozen states'' where all spins are dynamically inactive.
We aim to construct one-dimensional Krylov subspaces for generation $g > 1$ by utilizing the frozen states from generation one.
In the following, we consider the Vicsek fractal lattice of generation $g$ to consist of $5^{g - 1}$ first-generation fractal lattices.
We also consider two first-generation fractals to be nearest neighbors if they contain spins that are nearest neighbors.
A first-generation fractal has one, two, or four other first-generation fractals as nearest neighbors as illustrated in Fig.~\ref{fig:Vicsek-fractal-generation-3-colored}.
In particular, there are $5^{g - 2}$ first-generation fractals with four nearest neighbors.
On these first-generation fractals, we choose all sites to be spin-down.
The remaining $4\cdot 5^{g - 2}$ first-generation fractals have either one or two nearest neighbors, and they allow at least three frozen configurations as illustrated in Fig.~\ref{fig:Vicsek-fractal-generation-3-colored}.
We construct one-dimensional Krylov subspaces as the tensor product of frozen states on these first-generation fractals.
There are at least $3^{4 \cdot 5^{g - 2}} = 3^{4N/25}$ such one-dimensional Krylov subspaces where $N = 5^g$ is the total number of sites.
The number of Krylov subspaces, hence, grows exponentially with system size, and the model from Eq.~\eqref{eq:generalized-Hamiltonian} displays Hilbert space fragmentation on the Vicsek fractal lattice.

Next, we show that the fragmentation persists at a finite density of domain walls in the thermodynamic limit.
Similarly to the previous paragraph, we consider the Vicsek fractal lattice of generation $g$ to consist of $5^{g - 1}$ first-generation Vicsek fractal lattices.
The $4 \cdot 5^{g - 2}$ first-generation fractals with one or two nearest neighbors may be chosen from at least three frozen spin configurations as illustrated in Fig.~\ref{fig:Vicsek-fractal-generation-3-colored}.
One of these configurations does not introduce any domain walls since all sites are spin-down.
The remaining configurations have one spin-up which introduces two domain walls.
By choosing $m$ first-generation fractals as a frozen configuration with a single spin-up and all other first-generation fractals as spin-down, we construct a frozen state with $n_\text{dw} = 2m$ domain walls where $m = 0, 1, \ldots, 4\cdot 5^{g - 2}$.
The number of ways to choose the $m$ first-generation fractal lattices is given by the binomial coefficient ${4\cdot 5^{g - 2} \choose m}$.
Furthermore, for each first-generation fractal, there are at least two frozen configurations with a single spin-up. 
Therefore, the number of frozen states with $n_\text{dw} = 2m$ domain walls is at least
\begin{equation}
    {4N / 25 \choose n_\text{dw} / 2} 2^{n_\text{dw} / 2}.
    \label{eq:number_frozen_at_ndw}
\end{equation}
where $N = 5^g$ is the number of sites.
We aim to determine the number of frozen states at a fixed density of domain walls  $\rho_\text{dw} = n_\text{dw} / n_\text{dw}^\text{max}$ where $n_\text{dw}^\text{max}$ is the maximum number of domain walls on the Vicsek fractal lattice.
Before rewriting Eq.~\eqref{eq:number_frozen_at_ndw} in terms of the domain wall density $\rho_\text{dw}$, we determine the maximum number of domain walls $n_\text{dw}^\text{max}$.

The Vicsek fractal lattice is bipartite and the maximum number of domain walls is obtained by choosing all sites in one part as spin-down and all sites in the other part as spin-up.
In this case, all nearest neighbor edges are occupied by a domain wall.
Hence, the maximum number of domain walls is equal to the number of nearest neighbor edges.
In the following, we explicitly include the generation $g$ in the notation, i.e., $n_{\text{dw}, g}^\text{max}$ is the maximum number of domain walls on the Vicsek fractal lattice of generation $g$.
We determine the number of nearest neighbor edges recursively.
The Vicsek fractal lattice of generation $g + 1$ is constructed by connecting five Vicsek fractal lattices of generation $g$ as illustrated in Figs.~\ref{fig:Lattices}(a)-(c).
The number of nearest neighbor edges for generation $g + 1$ is, therefore, five times that for generation $g$ minus four corresponding to where the fractals of generation $g$ are connected.
The maximum number of domain walls follows the same relationship
\begin{equation}
    n_{\text{dw}, g + 1}^\text{max} = 5 n_{\text{dw}, g}^\text{max} - 4.
\end{equation}
The solution to the recursive relation with the initial condition $n_{\text{dw}, g = 1}^\text{max} = 8$ is given by
\begin{equation}
    n_{\text{dw}, g}^\text{max} = 7 \cdot 5^{g - 1} + 1
    \label{eq:n_dw^max}
\end{equation}

Inserting Eq.~\eqref{eq:n_dw^max} and $\rho_\text{dw} = n_\text{dw} / n_\text{dw}^\text{max}$ into Eq.~\eqref{eq:number_frozen_at_ndw}, we obtain a lower bound on the number of frozen states at domain wall density $\rho_\text{dw}$
\begin{equation}
    {4 N / 25 \choose \rho_\text{dw}(7 N / 5 + 1) / 2} 2^{\rho_ \text{dw} (7 N / 5 + 1) / 2}
    \label{eq:number_frozen_at_rho_dw}
\end{equation}
Recall that the number of domain walls for the considered frozen states is given by $n_\text{dw} = 2m$ with $m = 0, 1, \ldots, 4 \cdot 5^{g - 2}$.
Therefore, the expression in~\eqref{eq:number_frozen_at_rho_dw} is valid for all domain wall densities up to
\begin{align}
    \rho_\text{dw} &= \frac{8 \cdot 5^{g - 2}}{7 \cdot 5^{g - 1} + 1}
    \xrightarrow[g \to \infty]{} \frac{8}{35}.
    \label{eq:limit_rho}
\end{align}
Equations~\eqref{eq:number_frozen_at_rho_dw} and~\eqref{eq:limit_rho} demonstrate that the fragmentation persists at all domain wall densities $\rho_\text{dw} \leq 8/35$ in the thermodynamic limit since the number of Krylov subspaces scales exponentially with system size at these domain wall densities.

\section{Estimation of sector sizes and Krylov subspace dimensions}
\label{appendix:monte_carlo_methods}
Let $|s_1 \ldots s_N \rangle$ with $s_i \in \{-1/2, 1/2\}$ be a simultaneous eigenket of the Pauli $z$-operators $\{\sigma_{\bm r}^z\}$.
We consider the indicator function $\mathds 1_{n_\text{dw}}$ which signals whether $|s_1 \ldots s_N \rangle$ belongs to the symmetry sector with $n_\text{dw}$ domain walls
\begin{equation}
    \mathds 1_{n_\text{dw}}(|s_1 \ldots s_N \rangle) =
    \begin{cases}
        1, & \text{if $|s_1 \ldots s_N \rangle$ has } n_\text{dw} \\
        & \text{domain walls,} \\[1mm]
        0, & \text{otherwise.}
    \end{cases}
\end{equation}
The dimension of the symmetry sector with $n_\text{dw}$ domain walls is given by
\begin{equation}
    D_{n_\text{dw}} = \sum_{s_1, \ldots, s_N} \mathds 1_{n_\text{dw}}(|s_1 \ldots s_N \rangle)
    \label{eq:exact-sector-size}
\end{equation}
where the sum covers all $2^N$ possible configurations of the $N$ spins.
Equation~\eqref{eq:exact-sector-size} is impractical to evaluate numerically for large system sizes since the sum contains an exponential number of terms.
We circumvent this problem by employing Monte Carlo importance sampling.
We randomly draw $N_\text{MC} \ll 2^N$ product states $\{|\psi_i\rangle\}_{i=1}^{N_\text{MC}}$ that are simultaneous eigenkets of the Pauli $z$-operators.
Each product state is drawn independently from the probability distribution $P_p$ parameterized by $p \in [0, 1]$
\begin{equation}
    P_p(|s_1 \ldots s_N\rangle) = \prod_{i=1}^{N} p^{1/2 + s_i}(1-p)^{1/2 - s_i}.
\end{equation}
In other words, we draw each spin independently with probability $p$ of being spin-up and probability $1 - p$ of being spin-down.
The sector size is estimated by
\begin{equation}
    \widetilde{D}_{n_\text{dw}} = \frac{1}{N_\text{MC}} \sum_{i=1}^{N_\text{MC}} \frac{\mathds 1_{n_\text{dw}}(|\psi_i \rangle)}{P_p (| \psi_i \rangle)}.
    \label{eq:sector-size-estimate}
\end{equation}
and the estimate of the variance of $\widetilde{D}_\text{dw}$ is given by
\begin{equation}
    \sigma^2 = \frac{1}{N_\text{MC}(N_\text{MC} - 1)} \sum_{i = 1}^{N_\text{MC}} \left[\frac{\mathds 1_{n_\text{dw}}(|\psi_i\rangle)}{P_p(|\psi_i\rangle)} - \widetilde{D}_{n_\text{dw}} \right] ^ 2.
    \label{eq:variance-sector-size-estimate}
\end{equation}
For each generation $g$ and symmetry sector $n_\text{dw}$, we sample from the probability distribution $P_p$ which minimizes the variance, i.e., we perform a grid search and choose the $p$ that yields the smallest variance.

After computing $\widetilde{D}_{n_\text{dw}}$, we estimate the dimension of the largest Krylov subspace in this sector.
The largest Krylov subspace is determined from an exhaustive search of the full sector.
We randomly draw a product state with $n_\text{dw}$ domain walls.
We determine in which Krylov subspace the state resides and record the dimension of this subspace.
We repeat this procedure until $\widetilde{D}_{n_\text{dw}} - \tilde d_{n_\text{dw}}^\text{max}$ different product states have been found where $\tilde d_{n_\text{dw}}^\text{max}$ is the size of the largest Krylov subspace found so far.
The dimension of the largest Krylov subspace is then estimated by $\tilde d_{n_\text{dw}}^\text{max}$ after the search finishes.

\section{Construction of an exponential number of Krylov subspaces on the two-dimensional lattice}
\label{appendix:two_dimensional_lattice_fragmentation}
\begin{figure}
    \centering
    \includegraphics[width=\linewidth]{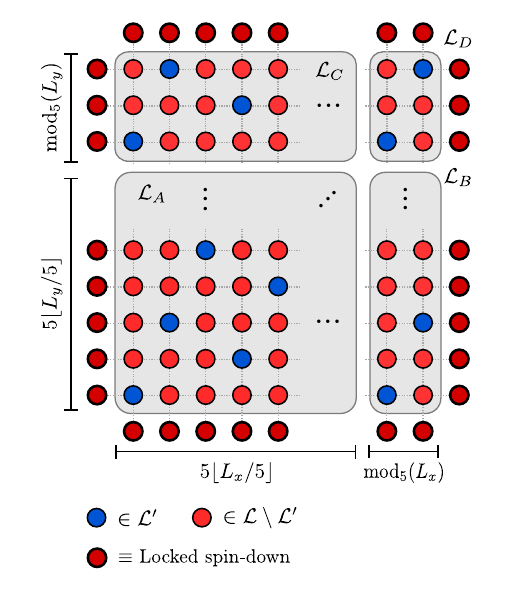}
    \caption{
        The two-dimensional lattice $\mathcal L$ of size $L_x \times L_y$ (light red and blue circles).
        The lattice is padded with extra sites along its boundary (dark red circles) to ensure all sites in the original lattice have an even number of nearest neighbors.
        The extra sites are dynamically inactive and we choose them to be spin-down.
        The lattice $\mathcal L$ does not contain the dynamically inactive sites.
        The sublattice $\mathcal L' \subset \mathcal L$ (blue circles) form a $5$-periodic pattern on $\mathcal L$ in the $x$ and $y$ directions.
        The lattice $\mathcal L$ is separated into four parts $\mathcal L_A$, $\mathcal L_B$, $\mathcal L_C$, and $\mathcal L_D$ (gray boxes).
        Part $\mathcal L_A$ is a rectangular lattice where both sides are a multiple of $5$.
       	Part $\mathcal L_B$ contains no sites when $\operatorname{mod}_5(L_x) = 0$ and $\mathcal L_C$ contains no sites when $\operatorname{mod}_5(L_y) = 0$.
       	When parts $\mathcal L_B$ and $\mathcal L_C$ are nonempty, they are rectangular lattices where one side is a multiple of $5$.
       	Part $\mathcal L_D$ is empty when $\operatorname{mod}_5(L_x) = 0$ or $\operatorname{mod}_5(L_y) = 0$.
        When part $\mathcal L_D$ is not empty it is a rectangular lattice where neither side is a multiple of $5$.
        }
    \label{fig:two_dimensional_lattice}
\end{figure}
We demonstrate that the model in Eq.~\eqref{eq:generalized-Hamiltonian} displays HSF on the two-dimensional lattice by constructing an exponential number of one-dimensional Krylov subspaces.
We consider the two-dimensional lattice of size $L_x \times L_y$ given by
\begin{equation}
    \mathcal L = \big\{(x, y) | x, y \in \mathds Z, \, 0 \leq x < L_x, \, 0 \leq y < L_y \big\}.
\end{equation}
We remark that the lattice $\mathcal L$ only contains the dynamically active sites and, hence, does not contain the extra sites introduced in Sec.~\ref{sec:model} to ensure all sites in the original lattice have an even number of nearest neighbors.
We also consider the sublattice $\mathcal L' \subset \mathcal L$ given by
\begin{equation}
    \begin{split}
        \mathcal L' = \big\{(n, 2n+5m) \vert n, m \in \mathds Z, \, 0 \leq n < L_x, \\
        0 \leq 2n+5m < L_y \big\}.
    \end{split}
\end{equation}
Figure~\ref{fig:two_dimensional_lattice} illustates the lattice $\mathcal L$ and the sublattice $\mathcal L'$.
The sites contained in $\mathcal L'$ form a pattern on the lattice $\mathcal L$ that repeats in the $x$ and $y$ directions with a period of $5$, i.e., if $(x, y) \in \mathcal L'$ and $0 \leq x \pm 5 < L_x$ then $(x \pm 5, y) \in \mathcal L'$ and, similarly, if $(x, y) \in \mathcal L'$ and $0 \leq y \pm 5 < L_y$ then $(x, y \pm 5) \in \mathcal L'$.

We aim to construct an exponential number of product states that are exact energy eigenstates of the Hamiltonian operator in Eq.~\eqref{eq:generalized-Hamiltonian}.
We consider product states of the form $\lvert s_1 s_2 \ldots s_{L_x L_y} \rangle$ where $s_i \in \{-1/2, 1/2\}$ is the spin along the $z$-direction on site $i$.
These product states are eigenstates of Eqs.~\eqref{eq:generalized-Hamilonian-magnetic-field-interaction} and \eqref{eq:generalized-Hamiltonian-nearest-neighbor-interaction} for all values of the spins $\{s_i\}_{i = 1}^{L_xL_y}$.
We ensure the product states are eigenstates of the full Hamiltonian operator by choosing the spins such that the kinetic term in Eq.~\eqref{eq:generalized-Hamiltonian-kinetic-term} annihilates the product states. 
We take all sites in $\mathcal L \setminus \mathcal L'$ to be spin-down where $\mathcal L \setminus \mathcal L'$ is the set difference, i.e., the sites contained in $\mathcal L$ but not in $\mathcal L'$.
We choose the sites in $\mathcal L'$ freely as either spin-down or spin-up.
Recall that the kinetic term in Eq.~\eqref{eq:generalized-Hamiltonian-kinetic-term} flips a spin if the sum of the magnetization of its nearest neighbors is zero.
When choosing the spins as described above, the sum of the magnetization of the nearest neighbors is either $-2$ or $-1$ for all sites.
Therefore, all product states of this form are eigenstates of the full Hamiltonian operator.
Since the sites in $\mathcal L'$ may be chosen as either spin-down or spin-up, the number of such product states is given by $2^{\lvert \mathcal L' \rvert}$ where $\lvert \mathcal L' \rvert$ is the number of sites in $\mathcal L'$.

We proceed by determining a lower bound on the number of sites in $\mathcal L'$.
First, we introduce some notation.
Let $\operatorname{mod}_5(L)$ be $L$ modulo $5$, i.e., the remainder after dividing $L$ by $5$.
Furthermore, let $\lfloor \cdot \rfloor$ be the function that rounds down to the nearest integer and notice that the expression $5 \lfloor L / 5 \rfloor$ rounds $L$ down to the nearest multiple of $5$.
We separate the lattice $\mathcal L$ into four parts as illustrated in Fig.~\ref{fig:two_dimensional_lattice}
\begin{align}
    \mathcal L_A &= \big\{(x, y) \in \mathcal L \mid x < 5 \lfloor L_x / 5 \rfloor, \, y < 5 \lfloor L_y / 5 \rfloor \big\}, \\
    \mathcal L_B &= \big\{(x, y) \in \mathcal L \mid 5 \lfloor L_x / 5 \rfloor \leq x, \, y < 5 \lfloor L_y / 5 \rfloor \big\}, \\
    \mathcal L_C &= \big\{(x, y) \in \mathcal L \mid x < 5 \lfloor L_x / 5 \rfloor, \, 5 \lfloor L_y / 5 \rfloor \leq y \big\}, \\
    \mathcal L_D &= \big\{(x, y) \in \mathcal L \mid 5 \lfloor L_x / 5 \rfloor \leq x, \, 5 \lfloor L_y / 5 \rfloor \leq y \big\}.
\end{align}
We also define the sites from $\mathcal L'$ in each of these parts by $\mathcal L_X' = \mathcal L' \cap \mathcal L_X$ with $X \in \{A, B, C, D\}$.
The number of sites in $\mathcal L'$ is given by 
\begin{equation}
    \lvert \mathcal L' \rvert = \lvert \mathcal L_A' \rvert + \lvert \mathcal L_B' \rvert + \lvert \mathcal L_C' \rvert + \lvert \mathcal L_D' \rvert.
    \label{eq:L^p}
\end{equation}
Part $\mathcal L_A$ represents a rectangular lattice with sides $5\lfloor L_x / 5 \rfloor$ and $5 \lfloor L_y / 5 \rfloor$ where both side lengths are a multiple of $5$.
Recall that the sites in $\mathcal L'$ form a pattern on the lattice $\mathcal L$ with period $5$ along the $x$ and $y$ directions.
Therefore, any horizontal sequence of five sites in $\mathcal L$, i.e., $(x, y), (x + 1, y), \ldots (x + 4, y) \in \mathcal L$, contains exactly one site in $\mathcal L'$.
Similarly, any vertical sequence of five sites in $\mathcal L$ also contain exactly one site in $\mathcal L'$.
Hence, the number of sites in $\mathcal L_A'$ is given by
\begin{equation}
    \lvert \mathcal L_A' \rvert = 5 \left\lfloor \frac{L_x}{5} \right\rfloor \left\lfloor \frac{L_y}{5} \right\rfloor.
    \label{eq:L_A^p}
\end{equation}
Part $\mathcal L_B$ is a rectangular lattice with side lengths $\operatorname{mod}_5(L_x)$ and $5 \lfloor L_y/5 \rfloor$.
Since the length of one side is a multiple of $5$, the number of sites in $\mathcal L'_B$ is given by
\begin{equation}
    \lvert \mathcal L_B' \rvert = \operatorname{mod}_5(L_x) \left\lfloor \frac{L_y}{5} \right\rfloor.
    \label{eq:L_B^p}
\end{equation}
Finally, part $\mathcal L_C$ forms a rectangle with side lengths $\operatorname{mod}_5(L_y)$ and $5 \lfloor L_x/5 \rfloor$.
Again, one side length is a multiple of $5$ and the number of sites in $\mathcal L_C'$ is given by
\begin{equation}
    \lvert \mathcal L_C' \rvert = \left\lfloor \frac{L_x}{5} \right\rfloor \operatorname{mod}_5(L_y).
    \label{eq:L_C^p}
\end{equation}
Inserting Eqs.~\eqref{eq:L_A^p}-\eqref{eq:L_C^p} into Eq.~\eqref{eq:L^p}, we find
\begin{equation}
        \lvert \mathcal L' \rvert = \frac{L_x L_y}{5} - \frac{1}{5} \operatorname{mod}_5(L_x) \operatorname{mod}_5(L_y) + \lvert \mathcal L_D' \rvert
\end{equation}
where we have utilized that $5 \lfloor L / 5 \rfloor = L - \operatorname{mod}_5(L)$.
We consider all possible values of $\operatorname{mod}_5(L_x), \, \operatorname{mod}_5(L_y) \in \{0, 1, 2, 3, 4\}$ and find that the expression $- \operatorname{mod}_5(L_x) \operatorname{mod}_5(L_y) / 5 + \lvert \mathcal L_D' \rvert$ is always larger than or equal to $-2/5$.
Hence, we find the lower bound on the number of sites in $\mathcal L'$ given by
\begin{equation}
    \lvert \mathcal L' \rvert \geq \frac{L_x L_y - 2}{5}
    \label{eq:Lp_lower_bound}
\end{equation}
In total, we find that the number of sites in $\mathcal L'$ scales linearly with the system size.
Therefore, the number of one-dimensional Krylov subspaces is larger than $2^{(L_x L_y  - 2) / 5}$ and, hence, grows exponentially with system size.
In summary, the model from Eq.~\eqref{eq:generalized-Hamiltonian} displays HSF on the two-dimensional lattice.

We show that the fragmentation persists at a finite density of domain walls in the thermodynamic limit.
We follow the same approach as in Appendix~\ref{appendix:Hilbert-space-fragmentation-proof} and utilize the frozen configurations constructed in the previous paragraph.
We consider product states where all sites are spin-down except the sites in $\mathcal L'$ which we choose as either spin-down or spin-up.
By choosing $m$ sites in $\mathcal L'$ as spin-up, we construct a frozen state with $n_\text{dw} = 4m$ domain walls where $m = 0, 1, \ldots, \lvert \mathcal L' \rvert$.
The binomial coefficient ${\lvert \mathcal L' \rvert \choose m}$ counts the number of ways of choosing $m$ sites from $\mathcal L'$.
Using the inequality~\eqref{eq:Lp_lower_bound}, we obtain a lower bound on the number of frozen states with $n_\text{dw}$ domain walls
\begin{equation}
    {(N - 2)/5 \choose n_\text{dw} / 4}
    \label{eq:number_frozen_at_ndw_2d}
\end{equation}
where $N = L_x L_y$ is the number of sites.
Before rewriting this expression in terms of the domain wall density $\rho_\text{dw} = n_\text{dw} / n_\text{dw}^\text{max}$, we determine the maximum number of domain walls $n_\text{dw}^\text{max}$ on the two-dimensional lattice.

The maximum number of domain walls is obtained by choosing every other site as spin-up and the remaining sites as spin-down. 
Concretely, we choose the lattice sites $\widetilde {\mathcal L} = \{(x, x + 2 \ell) \in \mathcal L \mid x ,\ell \in \mathds Z \}$ as spin-up and the remaining sites as spin-down.
Recall that the two-dimensional lattice is padded with extra sites as illustrated in Fig.~\ref{fig:two_dimensional_lattice} and we refer to these sites as the dynamically inactive sites.
For the spin configuration where all sites in $\widetilde {\mathcal L}$ are spin-up and the remaining sites are spin-down, the number of domain walls between dynamically active sites is given by $2L_x L_y - L_x - L_y$.
The number of domain walls between the dynamically active sites and the dynamically inactive sites depends on the size of the lattice.
If $L_x$ and $L_y$ are odd, there are $L_x + L_y + 2$ domain walls between these sites.
Otherwise, there are $L_x + L_y$ domain walls.
We cover both scenarios by writing the number of domain walls between the dynamically active sites and the dynamically inactive sites according to $L_x + L_y + 2  \operatorname{mod}_2(L_x)\operatorname{mod}_2(L_y)$.
We use the notation $\mu = \operatorname{mod}_2(L_x)\operatorname{mod}_2(L_y)$ in the following to keep the expressions concise.
The maximum number of domain walls is given by 
\begin{equation}
    n_\text{dw}^\text{max} = 2(N + \mu).
    \label{eq:n_dw^max_2d}
\end{equation}

We determine a lower bound on the number of frozen states at domain wall density $\rho_\text{dw}$ by inserting the expression $\rho_\text{dw} = n_\text{dw} / n_\text{dw}^\text{max}$ and Eq.~\eqref{eq:n_dw^max_2d} into Eq.~\eqref{eq:number_frozen_at_ndw_2d} and using Stirling's approximation for the factorial $x! \approx \sqrt{2 \pi x} x^x \exp(-x)$ which is valid in the limit $x \to \infty$
\begin{align}
    & {(N - 2)/5 \choose \rho_\text{dw}(N + \mu) / 2} \nonumber \\
    \xrightarrow[N \to \infty]{} & 
    \frac{1}{\sqrt{2 \pi N} \sqrt[10]{5}}
    \left( \frac{2}{\rho_\text{dw}}\right)^{\frac{1 + \mu \rho_\text{dw}}{2}}
    \left( \frac{10}{2 - 5 \rho_\text{dw}} \right)^{\frac{1 - 5 \mu \rho_\text{dw}}{10}}
    \nonumber \\
    & \times \left[ 
    \frac{1}{\sqrt[5]{5}}
    \left( \frac{2}{\rho_\text{dw}} \right)^{\frac{\rho_\text{dw}}{2}} 
    \left( \frac{10}{2 - 5 \rho_\text{dw}} \right)^{\frac{2 - 5 \rho_\text{dw}}{10}}
    \right]^{N}.
    \label{eq:number_frozen_at_rho_dw_2d}
\end{align}
For the domain wall densities $\rho_\text{dw} \in (0, 2 / 5)$, the expression in the square brackets in the third line of Eq.~\eqref{eq:number_frozen_at_rho_dw_2d} lies in the interval $(1, \sqrt[5]{2})$.
Hence, the number of frozen states at a fixed domain wall density scales exponentially with system size for these domain wall densities.
The number of domain walls of the considered frozen states is given by $n_\text{dw} = 4m$ with $m = 0, 1, \ldots, \lvert \mathcal L' \rvert$.
Using the inequality~\eqref{eq:Lp_lower_bound}, we find that the frozen states exist at least up to domain wall density
\begin{equation}
        \rho_\text{dw} = \frac{4 \lvert {\mathcal L}' \rvert }{n_\text{dw}^\text{max}} \geq \frac{2}{5} \frac{N - 2}{N + \mu}
        \xrightarrow[N \to \infty]{} \frac{2}{5}
        \label{eq:limit_rho_2d}
\end{equation}
in the thermodynamic limit.
Equations~\eqref{eq:number_frozen_at_rho_dw_2d}-\eqref{eq:limit_rho_2d} show that the fragmentation persists in the thermodynamic limit at a finite density of domain walls up to at least $\rho_\text{dw} = 2 / 5$.
\section{Perturbing the Hamiltonian with a random matrix}\label{appendix:perturbation}
\begin{figure}
    \centering
    \includegraphics[width=\linewidth]{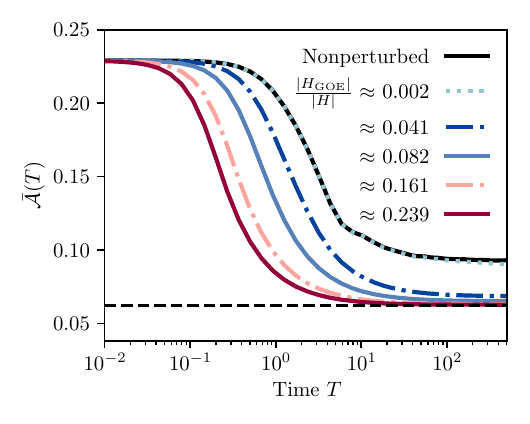}
    \caption{
    The time-averaged autocorrelation function on the Vicsek fractal lattice for the perturbed Hamiltonian from Eq.~\eqref{eq:perturbedhamiltonian}.
    The size of the perturbation is characterized by the ratio between the Frobenius norm of the perturbation matrix $|H_{\mathrm{GOE}}|$ to that of the nonperturbed Hamiltonian $|H|$.
    The dashed line displays the Mazur bound obtained from the projection operators onto the Krylov subspaces.
    }
    \label{fig:autocorrelationswith perturbation}
\end{figure}
In Sec.~\ref{sec:autocorrelationfunctions}, we observe that the time-averaged autocorrelation function does not converge to the Mazur bound obtained from the projection operators onto the Krylov subspaces.
In this appendix, we demonstrate that the Mazur bound becomes tight when the Hamiltonian operator is perturbed.
We perturb the Hamiltonian operator by a block diagonal matrix whose blocks are drawn from the Gaussian orthogonal ensemble (GOE).
The GOE consists of symmetric matrices where the entries follow the normal distribution $\mathrm{N}(\mu, \sigma^2)$.
The diagonal entries have zero mean and variance $\sigma^2 = 2$, i.e., $G_{ii} \sim \mathrm{N}(0,2)$, while the off-diagonal entries have zero mean and unit variance, i.e., $G_{ij} \sim \mathrm{N}(0,1)$ for $i \neq j$.
The entries are independent up to the symmetry requirement, i.e., $G_{ij} = G_{ji}$ \cite{Anderson_Guionnet_Zeitouni_2009}.
The block diagonal structure is chosen to overlap with that of the kinetic term of the Hamiltonian to conserve the HSF.
The Hamiltonian is given by
\begin{align}
    H' = H + H_\text{GOE} = H_\lambda + H_z + H_{zz} + H_\text{GOE},
    \label{eq:perturbedhamiltonian}
\end{align}
where the terms $H_\lambda$, $H_z$ and $H_{zz}$ are defined in Eq.~\eqref{eq:generalized-Hamiltonian-terms} and  $H_\text{GOE} = \epsilon G$, where $G$ is a block diagonal matrix of GOE matrices and $\epsilon \in \mathbb R$ is a strength parameter.
We characterize the strength of the perturbation by the ratio of the Frobenius norms.
Figure~\ref{fig:autocorrelationswith perturbation} shows that the time-averaged autocorrelation function for the chosen spin operator converges to the Mazur bound when the Hamiltonian is perturbed.

\bibliography{bibliography}

\end{document}